\documentclass{aa}  

\usepackage{graphicx}
\usepackage{txfonts}
\usepackage[usenames]{xcolor}

\newcommand{\Ks}{\mathrm{K_s}}
\newcommand{\J}{\mathrm{J}}
\DeclareUnicodeCharacter{2212}{-}

\begin{document}

   \title{Mapping the stellar age of the Milky Way bulge with the VVV. 
   \thanks{Based on observations 
taken within the ESO VISTA Public Survey VVV, Program ID 179.B-2002 (PI: Minniti, Lucas)}
   \thanks{The result is publicly available at http://basti-iac.oa-teramo.inaf.it/vvvexmap/}
    \thanks{Additionally, the map is also available in electronic form
at the CDS via anonymous ftp to cdsarc.u-strasbg.fr (130.79.128.5)
or via http://cdsweb.u-strasbg.fr/cgi-bin/qcat?J/A+A/}}

   \subtitle{ III. High resolution reddening map.}

  \author{F. Surot,
          \inst{1,2}
    E. Valenti,
         \inst{3,4}
    O. A. Gonzalez,
         \inst{5}
    M. Zoccali,
          \inst{6,7}
    E. S{\"o}kmen,
         \inst{1,2}
    S. L. Hidalgo,
         \inst{1,2}
   D. Minniti,
         \inst{7,8,9}    
         }
\institute{Instituto de Astrof\'{i}sica de Canarias, E-38205, La Laguna, Tenerife, Spain\\
\email{frsurot@uc.cl}
\and
Departamento de Astrof\'{i}sica, Universidad de La Laguna, E-38205, La Laguna Tenerife, Spain
\and
European Southern Observatory, Karl Schwarzschild-Stra\ss e 2, D-85748 Garching 
bei M\"{u}nchen, Germany. 
\and
Excellence Cluster ORIGINS, Boltzmann-Stra\ss e 2, D-85748 Garching bei M\"{u}nchen, Germany
\and
UK Astronomy Technology Centre, Royal Observatory, Blackford Hill, Edinburgh, EH9 3HJ, UK
\and
Instituto de Astrof\'{i}sica, Pontificia Universidad Cat\'{o}lica de Chile, Av. Vicu\~{n}a Mackenna 4860, Santiago , Chile.
\and
Millennium Institute of Astrophysics, Av. Vicu\~{n}a Mackenna 4860, 782-0436 Macul, Santiago, Chile.
\and
Departamento de Ciencias F\'{i}sicas, Universidad Andr\'{e}s Bello, Rep\'{u}blica 220, Santiago, Chile
\and
Vatican Observatory, V00120 Vatican City State, Italy
             }

   \date{}

 
  \abstract
   {The detailed study of the Galactic bulge stellar population necessarily requires an accurate representation of the interstellar extinction particularly toward the Galactic plane and center, where the severe and differential reddening is expected to vary on sub-arcmin scales.  Although recent infrared surveys have addressed this problem by providing extinction maps across the whole Galactic bulge area, dereddened color-magnitude diagrams near the plane and center appear systematically undercorrected, suggesting the need for higher resolutions. These undercorrections affect any stellar study sensitive to color (e.g. star formation history analysis via color-magnitude diagram fitting), either making them inaccurate or limiting them to small low/stable extinction windows where this value is better constrained.}
   {We aim at providing a high-resolution (2\,arcmin to $\sim$\,10\,arcsec) color excess map for the VVV bulge area, in $\J-\Ks$ color.}
   {We use the MW-BULGE-PSFPHOT catalogs sampling $\sim$\,300\,deg$^2$ across the Galactic bulge ($|l| < 10^\circ$ and $-10^\circ < b < 5^\circ$) to isolate a sample of red clump and red giant branch stars, for which we calculate average $\J-\Ks$ color in a fine spatial grid in $(l, b)$ space.}
   {We obtain a E$(\J-\Ks)$ map spanning the VVV bulge area of roughly 300\,deg$^2$, with the equivalent to a resolution between $\sim$\,1\,arcmin for bulge outskirts ($l < −6^\circ$) to below 20\,arcsec within the central $|l| < 1^\circ$, and below 10\,arcsec for the innermost area ($|l| < 1^\circ$ and $|b| < 3^\circ$). The result is publicly available at http://basti-iac.oa-teramo.inaf.it/vvvexmap/}
   {}

   \keywords{Galaxy: structure -- Galaxy: bulge } \authorrunning{Surot et al.}  

   \maketitle
%

\section{Introduction}
\label{sec:intro}

Despite the large efforts undertaken in the last decade to construct reddening maps in the Galactic bulge, interstellar extinction remains as one of the most uncertain parameters when studying the stellar populations of the inner regions of the Milky Way. A combination of systematics in the construction of reddening and extinction maps, as well as the uncertainties in the extinction law used to obtain reddening-free magnitudes can have considerable effects in the results based on both photometric and spectroscopic observations of the central regions of the Galaxy.

The extinction law has been shown to be non-standard and variable in the regions closer to the Galactic plane \citep[see][for a detailed review]{nataf+16}, thus results can differ depending on the bands used for each study and the adopted extinction law. On the other hand, the main limitation for the applicability of reddening maps on studies of the Galactic bulge comes from their spatial resolution. Reddening variations on very small scales, down to a few arcseconds, can be seen in maps of specific windows \citep[see for example][]{gosling+09}. The residual differential reddening within the resolution element of a reddening map can strongly affect the extinction correction of stellar populations or the characterization of the extinction law. As a result, and with the increasing number of photometric and spectroscopic surveys covering the bulge, special attention has been given to improve the spatial scales in which reddening is calculated without compromise from sensitivity. However, wide-field reddening and extinction maps of the Galactic bulge in sub-arcmin scales are not yet available.

With the advent of infrared (IR) and near-IR photometry, the studies of the Galactic bulge have advanced during the last decade. Large surveys (DENIS - \citealt{denis}; 2MASS - \citealt{2mass}; UKIDSS - \citealt{ukidss}; VVV(X) - \citealt{vvvx}; GLIMPSE - \citealt{glimpse}) provided a framework necessary to take the extinction studies to regions of higher extinction, thus closer to the Galactic plane at considerably higher spatial resolution than previously available. For example, \citet{schultheis99} used DENIS photometry to produce color-magnitude diagrams (CMDs) in 2\,arcmin windows and compare the location of M giants with theoretical isochrones of red giant branch (RGB) and asymptotic giant branch (AGB) stellar populations. This comparison was used in \citet{schultheis99} to estimate reddening in each of these windows to produce the first high-resolution extinction map of the Galactic plane, including regions with extinction as large as 35 magnitudes in the optical. However, this study is limited by the sensitivity of DENIS in these regions, in particular $\J$-band. \citet{dutra+03} later produced a map applying the same technique to 2MASS data, but therefore covering a considerably larger area. In this case, the limitation of the map is coming from the spatial resolution of 2MASS that is not sufficient to avoid the crowding effects in the large density, innermost regions of the Galaxy.

This situation improved dramatically thanks to the spatial resolution and depth of near-IR of the VISTA Variables in the V\'{i}a L\'{a}ctea (VVV) survey \citep{minniti_vvv} which allowed different teams to map red-clump (RC) stars (as previously done with optical photometry in the outer bulge) all the way down to the Galactic plane. The first VVV extinction map was presented by \citet{oscar12} with a spatial resolution ranging from 6\,arcmin in the outer bulge to 2\,arcmin near the Galactic plane. Extinction maps with the same spatial resolution were constructed using GLIMPSE by \citet{schultheis09}, following the same isochrone-fitting technique from \citet{schultheis99}. Because in the case of the VVV extinction maps, the spatial resolution of the maps was determined by the number of RC stars in the resolution element, the achieved resolution was recently improved to 1\,arcmin resolution thanks to the construction PSF-catalogs \citep{alonso_vvv} with deeper and more complete photometry down to the RC magnitude level across the entire bulge region \citep{oscar18}, achieving the highest coverage/resolution ratio for a 2D extinction map of the Galactic bulge region so far.

Extinction maps are also useful to identify windows of low extinction at low latitudes within the plane, that are strategic to see deeply through the Milky Way. A classic example would be Baade’s Window, that has traditionally been important for the studies of the stellar populations in the Galactic bulge. More recently, the VVV survey has allowed the identification of a couple more windows with relatively low and uniform extinction in the Galactic plane \citep{minnitiwin,saitowin}. These windows are VVV WIN 1713-3939 located at Galactic coordinates $(l, b) = (347.4^\circ, -0.4^\circ)$, with mean total near-IR extinction $\mathrm{A}_\Ks = 0.46$\,mag, and VVV WIN 1733-3349 located at $(l, b) = (-5.2^\circ, -0.3^\circ)$, with mean extinction $\mathrm{A}_\Ks = 0.61$\,mag.

However, and despite the progress made in this regard, the Galactic center and plane remain a challenge to observe, in great part due to the high reddening variations in these highly-extincted areas, which are unaccounted for even by arcmin resolutions in recent extinction maps. If we are to study the Galactic bulge stellar populations near the center or plane \citep[e.g.,][]{methodpaper} much higher resolutions are needed to either correct for extinction or emulate it in theoretical frameworks.

In this work, we present a new high resolution projected reddening map, in terms of $\mathrm{E}(\J-\Ks)$ in the VISTA system, reaching unprecedented sub-arcmin resolution near and on the Galactic plane and covering the whole VVV bulge area ($\sim$\,320\,deg$^2$, $|l| < 10^\circ$ and $-10^\circ < b < 5^\circ$). In the coming sections we briefly detail the data used in this work (Section~\ref{sec:data}), the methodology used to derive the map (Section~\ref{sec:method}), some comparison with existing reddening maps in the literature (Section~\ref{sec:comparison}), and its caveats (Section~\ref{sec:caveats}). We finalize with a summary and short discussion (Section~\ref{sec:end}).


\section{Dataset}
\label{sec:data}

This work fully exploits the accuracy and depth of the MW-BULGE-PSPHOT compilation based on VVV images taken with VIRCAM@VISTA/Paranal Observatory, and publicly available from the ESO Science Archive Portal\footnote{http://archive.eso.org/scienceportal/home}.
The detailed description of the data reduction, completeness study and catalogs construction has been presented in \cite{photopaper}.
What follows is, therefore, only a brief summary of the main dataset properties.

The VVV bulge footprint is a mosaic of 196 tiles covering a continuous total area of $\sim$\,320\,deg$^2$ around the Galactic center, with about 10\% overlap among adjacent tiles.
Globally, the MW-BULGE-PSFPHOT contains nearly 600 million stars detected across the bulge area surveyed by the VVV; however, it is arranged in a set of 196 catalogs obtained by performing PSF-fitting photometry of multi-epoch $\J\Ks$ images.
Each catalog provides a homogeneous sky coverage of $1.5^\circ\times1.2^\circ$, and contains all stars detected within a given tile through both passband filters, $\J$ and $\Ks$.
In addition, each star detection is coupled with its corresponding photometric completeness as derived through extensive artificial stars experiments.

With limiting magnitudes $\Ks\sim20$ and $\J\sim21$, the MW-BULGE-PSFPHOT photometric compilation allows to study the evolved and un-evolved stellar population of the Milky Way bulge over most of its extension.
The RC population is properly sampled with a photometric completeness ranging from nearly 100\% to 70\% throughout the VVV bulge area.
Exception are some of the innermost fields close to the Galactic center where the completeness drops to $\sim$50\%.
In addition, the photometry is accurate and deep enough to sample the old main-sequence (MS) turnoff across the whole outer bulge region (i.e., $|b|\gtrsim3.5^\circ$) with over 50\% completeness.

\begin{figure*}[t]
        \centering
        \includegraphics[width=0.49\hsize]{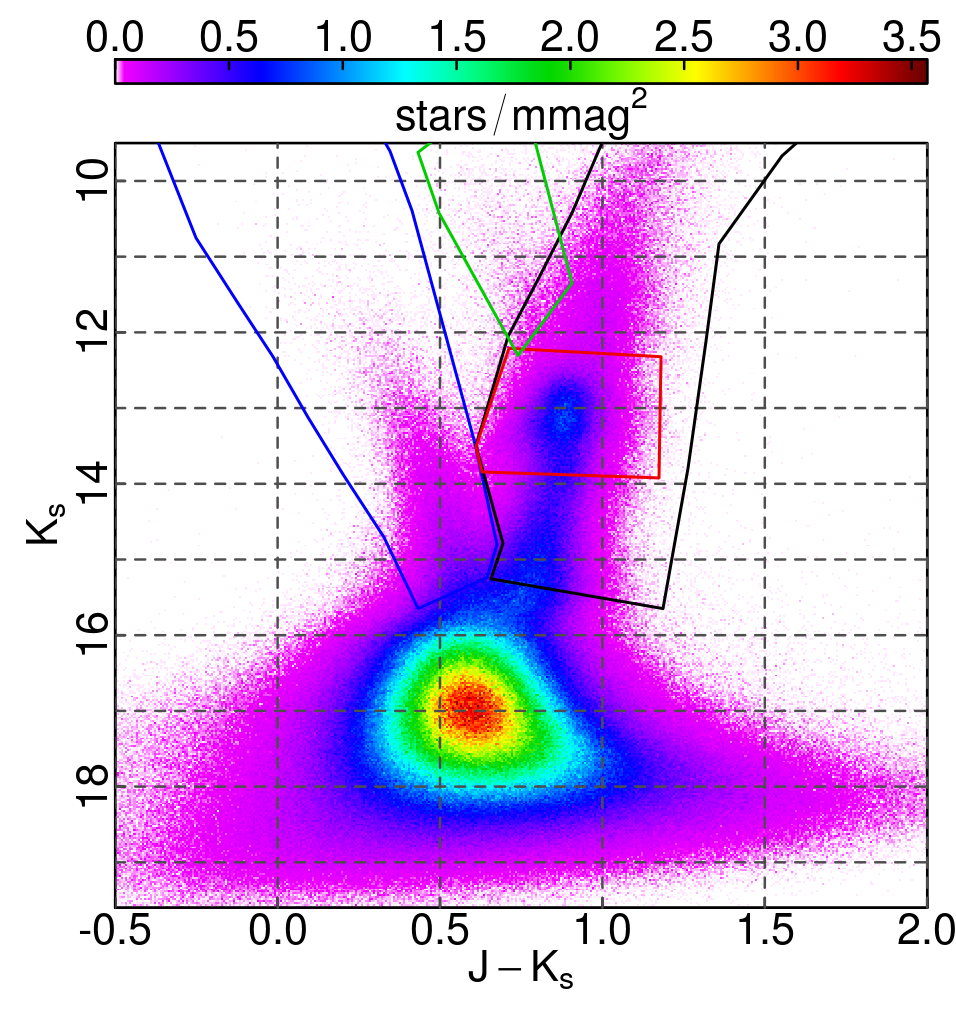}
        \includegraphics[width=0.49\hsize]{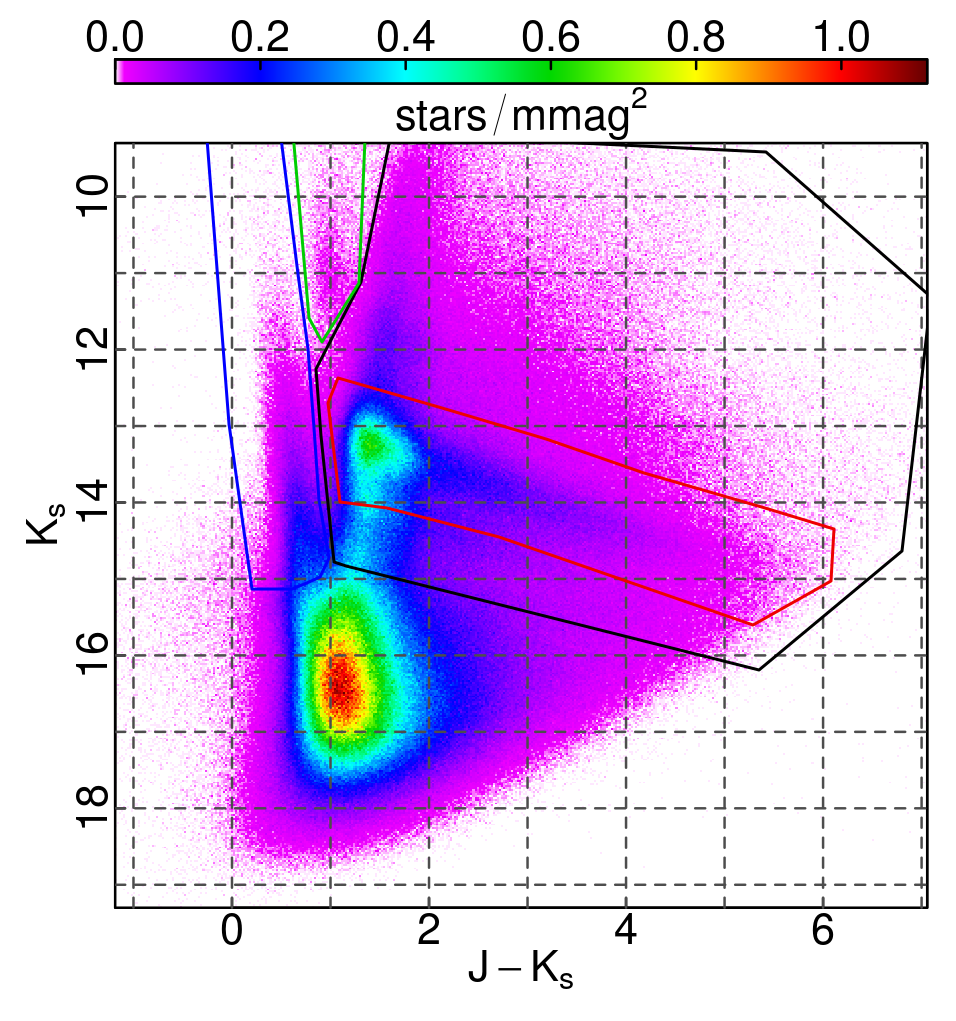}
        \caption{Hess diagram of tiles b278 (left panel; $l \approx 1^\circ$ and $b \approx -4^\circ$, with $0.12 \lesssim E(\J-\Ks) \lesssim 0.67$) and b320 (right panel; $l \approx 1^\circ$ and $b \approx -1^\circ$, with $0.40 \lesssim E(\J-\Ks) \lesssim 5.30$). Different  polygons are used to roughly highlight some of the sampled evolutionary sequences: the bright disk MS (blue polygon), the bulge RGB (black polygon), the RC (red polygon) and the blue plume of evolved disk sequence (green polygon).}
        \label{fig:samplecmd}
\end{figure*}

Because the entire photometric compilation is very diverse and extensive, as an example in Fig.~\ref{fig:samplecmd} we show the derived CMD of two selected fields located in regions characterized by very different crowding and reddening.
Tile b278 (Fig.~\ref{fig:samplecmd}, left panel) is roughly centered in the low reddening Baade's Window field ($l \approx 1^\circ$ and $b \approx -4^\circ$), while tile b320 is located on the Galactic plane ($l \approx 1^\circ$ and $b \approx -1^\circ$) where the extinction and differential reddening are much more severe.
In Fig.~\ref{fig:samplecmd}, different polygons are used to roughly highlight the main evolutionary sequences sampled by the photometric catalogs, namely: the bulge RC (red), the bulge RGB (black), the brightest portion of the disk MS (blue) and the blue plume of the evolved disk stars (green).
The differences in the amount of extinction and its distribution within the two fields are clearly seen when comparing the mean color and spread in color of the bulge RC and RGB in the CMDs.
On the other hand, the disk MS and blue plume do not show any remarkable or noticeable color spread between the fields, hence suggesting the extinction being confined within the bulge only, and not affecting the foreground disk.

\section{Methodology}
\label{sec:method}

Here we describe the method that has been applied to measure color excess using RGB and RC stars in all VVV tiles, hence producing 196 individual reddening maps of $610 \times 500$ pixels with a field-of-view of roughly 1.6\,$\mathrm{deg}^2$. In terms of figures and concrete examples, we will use a representative tile, namely b320 ($0.21^\circ <  l < 1.69^\circ$ and $-1.56^\circ < b < -0.35^\circ$), whose CMD is displayed in Fig.~\ref{fig:error}. This tile is especially useful as a template, because of the consistently prominent and compact (in $\Ks$) RC trail in color-magnitude it provides, extending up to $\J-\Ks \sim 5$, which is a desirable property when attempting to define a reddening law based on the RC color-magnitude correlation.

\subsection{Selection of RGB and RC stars}
\label{sec:selection}

To estimate the color excess E($\J-\Ks$) we take advantage of the relatively vertical sequence traced by the bulk of the RGB and RC population in the [$\Ks$ vs $(\J-\Ks)$] plane (see Fig.~\ref{fig:samplecmd}). To isolate the RGB and RC population from the rest of the sampled stars within a given tile, we applied a color-magnitude selection in the [$\J-\Ks$] vs. [$\Ks$] CMD. 

Manually and for each field, we selected all stars within a region whose color limits are set such as to include RGB and RC stars only, therefore excluding the bluer disk MS and blue plume (see blue and green boxes in Fig.~\ref{fig:samplecmd}), traced by the young (and foreground) disk stars.
This isolation is key, because the foreground disk population is likely affected by a different reddening than the background bulge population \citep[see][]{schultheis14}, and thus it must be excluded from the selected sample.

We have defined the bright and faint limits of the selection region to vary with color following the reddening vector while containing the bulk of the RC, as well as slightly brighter and fainter RGB stars, whenever this would not also include blue plume or evolved disk stars. Regarding the reddening vector, we derive it directly from the color-magnitude trend of the RC in tile b320, modeled following a similar method to that described in \citet{alonsogarcia+17}. The measured reddening vector is:

   \begin{align}
   	A_{\Ks} &= 0.422167\,\mathrm{E}(\J-\Ks) \\
   	A_{\J} &=  1.422167\,\mathrm{E}(\J-\Ks)
   \label{eq:redlaw}
   \end{align}

\begin{figure}
        \centering
        \includegraphics[width=0.99\hsize]{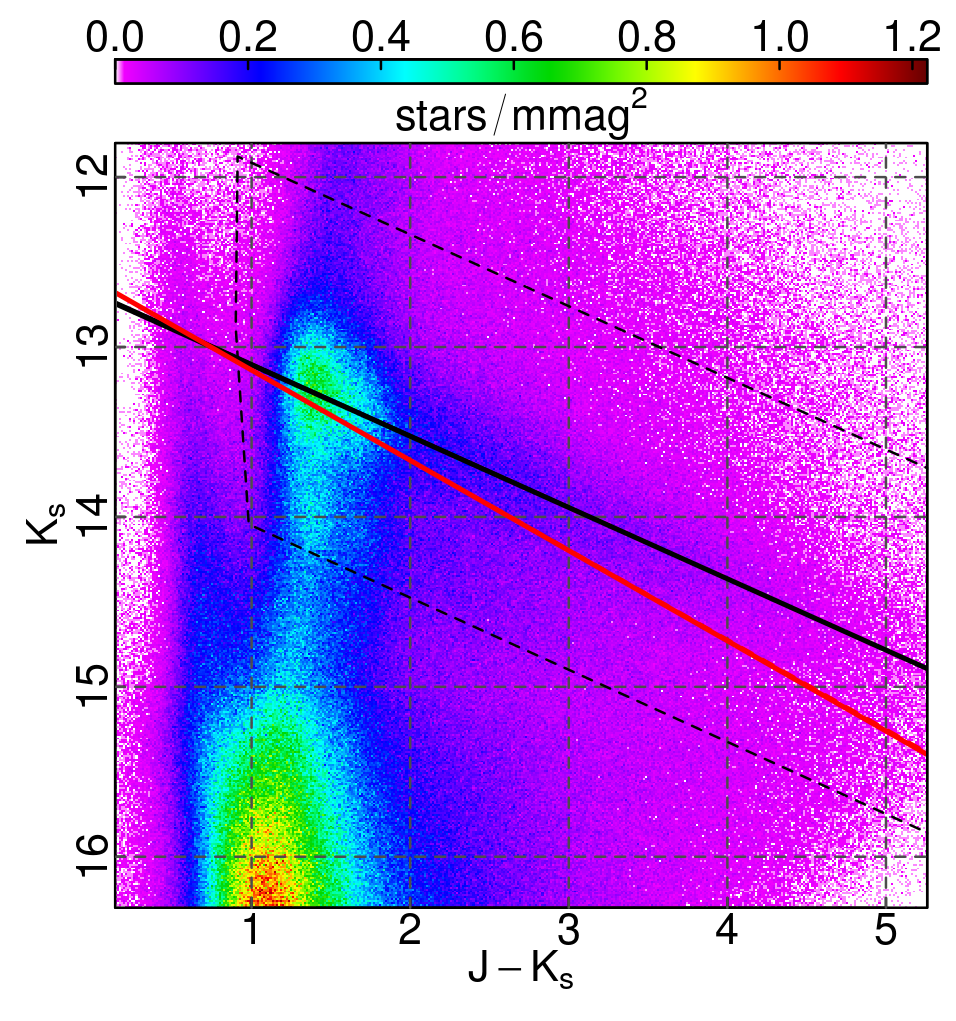}
        \caption{Zoomed-in Hess diagram of the selected field b320. The solid black line follows the derived reddening vector (see text), while the dashed black box defines the polygon/region used to select the stars in the sample for the color excess calculations. Additionally, the red solid line follows \citet{nishiyama+09} reddening law, and intersects the solid black line at $\left(\J-\Ks\right) = 0.635$, which is roughly the $\left(\J-\Ks\right)_0$ of the reddening corrected RC+RGB sequence at their peak overdensity in the CMD.}
        \label{fig:error}
\end{figure}   

In the sample field b320, the adoption of this empirical reddening vector produces a smaller dispersion in $\Ks$ for the stars in RC after correction ($\sigma_{\Ks} = 0.280$\,mag) than when using a fixed reddening law such as \citet{nishiyama+09} ($\sigma_{\Ks} = 0.299$\,mag). In Fig.~\ref{fig:error} we show the CMD of b320 with the reddening vector derived here, together with the CMD region/polygon used to define the stellar sample, and a comparison with the reddening law from \cite{nishiyama+09}. We will prefer this empirical law throughout this work.
After we define the selection polygon in color-magnitude space, we pick all the stars within it and save it to a sample stellar catalog of RC and RGB stars, for which we only keep position and color information (RC+RGB catalog).

\subsection{Mapping the color of the RGB and RC sequences}

We proceed by defining a grid in Galactic latitude ($b$) and longitude ($l$), which covers the area of the original field of view (in the case of VVV, roughly $1.5 \times 1.2$\,deg$^2$) with regularly spaced nodes; 610 in $l$ and 500 in $b$, defining 305,000 pixels/nodes in the grid. This grid will define the resolution of our map, with pixel sizes $<10\,\mathrm{arcsec}$. This scale is the limit at which we can trace line-of-sight variations in color of our sample RC+RGB stars.

For each node defined by the grid, we look at the closest 20 stars (a parameter henceforth referred as $NAB$) in the sample, based on their angular distances to the center of each pixel and filter out outliers, defined as stars whose colors are more than 3 median absolute deviation (MAD) from the ensemble median. The information of the remaining $N$ stars can be reduced to a set of pairs $\{(r_i,c_i)\}_{i=1}^N$, where $r_i$ is the angular distance of the $i$-th star to the pixel center and $c_i$ its $\J-\Ks$ color. With this we define a set of weights $w_i$ so that:

   \begin{align}
   	w_i &= 1-0.75 \left( \frac{r_i}{r_\mathrm{max}} \right)^2
   \label{eq:cweights}
   \end{align}

Where $r_\mathrm{max}$ is the maximum from $\{r_i\}_{i=1}^N$. The weights are normalized to sum 1. These weights are mostly arbitrary, designed so that tracers near the center of the node have close-to-maximal weights while the farthest star(s) still contribute to the calculations. For instance, if we had only 4 tracers near the center, while the other 16 lie at the border of maximum distance to the center, $r_\mathrm{max}$, then the central stars would have a contribution of about 50\% by weight.

Finally, with these weights, we define for each node four values. A weighted average of the color of the $N$ stars ($ee$), the weighted variance ($\sigma_{ee}^2$), and the weighted average and maximum distance of the $N$ stars to the center of the pixel ($\bar{r}$ and $r_{max}$, respectively).

The product of this algorithm is a cube that contains the $[l,b]$ maps for each of those quantities. The $ee$ values are the color excess tracers we need. Although $\sigma_{ee}/\sqrt{NAB}$ could be understood as an estimate of the uncertainty in $ee$, $\sigma_{ee}$ is truly tracing the width of the color sequence distribution of the stars used in the calculation. The $\bar{r}$ and $r_{max}$ values will trace the definition of the map, that is, the smallest size of the structures the map is able to trace.

The way these maps are constructed produces increasingly better definition the higher the stellar density of the field gets, which can get below 20\,arcsec near the Galactic plane and center. Additionally, it is important to note that we make the distinction between resolution and definition. The maps we recover here are actually an evaluation of a mosaic/tessellation defined by the data (i.e. stellar density of the RC+RGB sample) and the number of stars allowed in the calculation ($NAB$). This mosaic is similar to a Voronoi tessellation, but where each polygon defines the area closest to a unique combination of $NAB$ stars, which can be smaller in size than the distance between the considered stars themselves (see Appendix~\ref{ap:highervor}). This opens the possibility to arbitrarily increase the resolution (i.e. the number of nodes in the grid), which would in principle increase the fidelity of the evaluation, but will have no effect on the definition of the map, which is fixed by the stellar density of the RC+RGB sample and is, as a rule of thumb, the size of the smallest structures (e.g., clouds, filaments, etc) we can trace.

\subsection{Calibrated $\mathrm{E}(\J-\Ks)$ reddening maps}
\label{sec:calib}

Once we have applied this algorithm to all 196 tiles in the VVV, we proceed to calibrate the $ee$ values to useful E$(\J-\Ks)$.

For an internal calibration, we have taken a \textit{whack-a-mole} approach, where given a comparison metric, we correct/calibrate outliers until the whole ensemble is consistent. In practice, we have taken every tile $ee$ map and made a comparison with the $ee$ values from the maps of their direct neighbors. That is, given that each tile has a small overlap ($\sim$10\% total area) with its immediate neighbors, we can take the mean difference between the $ee$ values of a tile within these overlapping regions and the $ee$ values of the respective neighbors, and provide up to 4 $\Delta ee = \bar{ee}_\mathrm{tile}-\bar{ee}_\mathrm{neighbor}$ values (so only 3 for tiles at the border of the bulge, and 2 for the corner fields). Since each field has a particular selection window, we expect $\Delta ee$, or rather, the mean of the $\Delta ee$ values for each field ($\left<\Delta ee\right>$) to be non-zero, due to the dependence of the mean $\J-\Ks$ color of the RC+RGB sequence with this selection. In a perfectly consistent ensemble, it would be possible to find a set of zero-point corrections so that all the $\left<\Delta ee\right>$ values for every field would become 0, however we expect to have some residual non-trivial discrepancy from the photometry \citep[see][]{photopaper}, which in practice means that we can at best aim for a minimal or reasonable dispersion around 0. Fortunately, the concrete set of $\left<\Delta ee\right>$ values for our ensemble is already close to optimal, with only a few outliers (mostly near the center). Given the independent nature of each field, an easy way to achieve the calibration is to iteratively \textit{hammer down} (or \textit{up}) the strongest outlier with a zero-point correction to bring it to level with its neighbors, recalculate the $\left<\Delta ee\right>$ of the ensemble and calibrate the next strongest outlier after this correction, so that after the process is done, all the \textit{bumps} are flattened and the set is more or less consistent within some dispersion. In particular, we have defined as outlier any field whose $\left<\Delta ee\right>$ is outside of the $\pm 2.5\sigma$ of the ensemble. This choice produces a fast convergence, were only 32 tiles where ultimately affected, and where the final distribution of $\left<\Delta ee\right>$ is indistinguishable from a gaussian distribution with mean 0 and $\sigma = 0.007$\,mag.

For the absolute calibration, we have used color excess values from the map of \cite{oscar12} and the improved central area from \cite{oscar18} (a map referred to as G18 henceforth) as a reference, which is also based in VVV data and uses fits to the $(\J-\Ks)$ color distribution of RC stars in binned fields of view and obtains an E$(\J-\Ks)$ value by comparing with a reference RC color in Baade's Window. With respect to the original \cite{oscar12} version, G18 boasts better resolution along the Galactic plane, as well as solving a few glitches in the central $|l| \lesssim 9.5^\circ, b \lesssim 4.5^\circ$ area see Section~\ref{sec:comparison} and Fig.~\ref{fig:b320_triad}). The choice for G18 is motivated by the idea that a comparison between the calibrator and this work should have minimal sources of uncertainty, and the similarities listed above limit the effective difference between this work and G18 to (mostly) just resolution and photometric depth. Other maps in literature would introduce additional uncertainty by requiring a projection (in case of 3D maps), and/or a system conversion/reddening law (in case of different passbands). Also, G18 reports a resolution of 1\,arcmin in the best cases, which is the best available for its coverage. However, G18 is less sensitive to high-extinction areas than the map presented in this work. This, and the expected abrupt changes in extinction near the plane, mean that for stars near the Galactic center and plane, we necessarily will have some difference between our measurements and those in G18. For this reason, in order to obtain a straightforward zero-point calibration, we have used only areas with $|b| > 3^\circ$, which yielded an absolute calibration of $ee-0.605 = \mathrm{E}(\J-\Ks)$. The field-to-field internal calibration and this global comparison essentially puts the color excess values we obtain on roughly the same reference as the one used in G18. In particular, we can compare the VVV tile containing Baade's Window (b278, $l \approx 1^\circ, b \approx -4^\circ$) and correct RC stars for both extinction by G18 and the map presented here with the reddening law from (\ref{eq:redlaw}). In this case, the RC stars' color standard deviation (and MAD) after correcting with this work's map is 85\% (and 83\%) of the obtained from the literature extinction, while maintaining a difference in median color of only 0.004\,mag.

\section{Results}

\subsection{High spatial resolution map}

\begin{figure*}[ht]
\centering
\includegraphics[width=0.99\hsize, trim={7cm 0 6cm 0}]{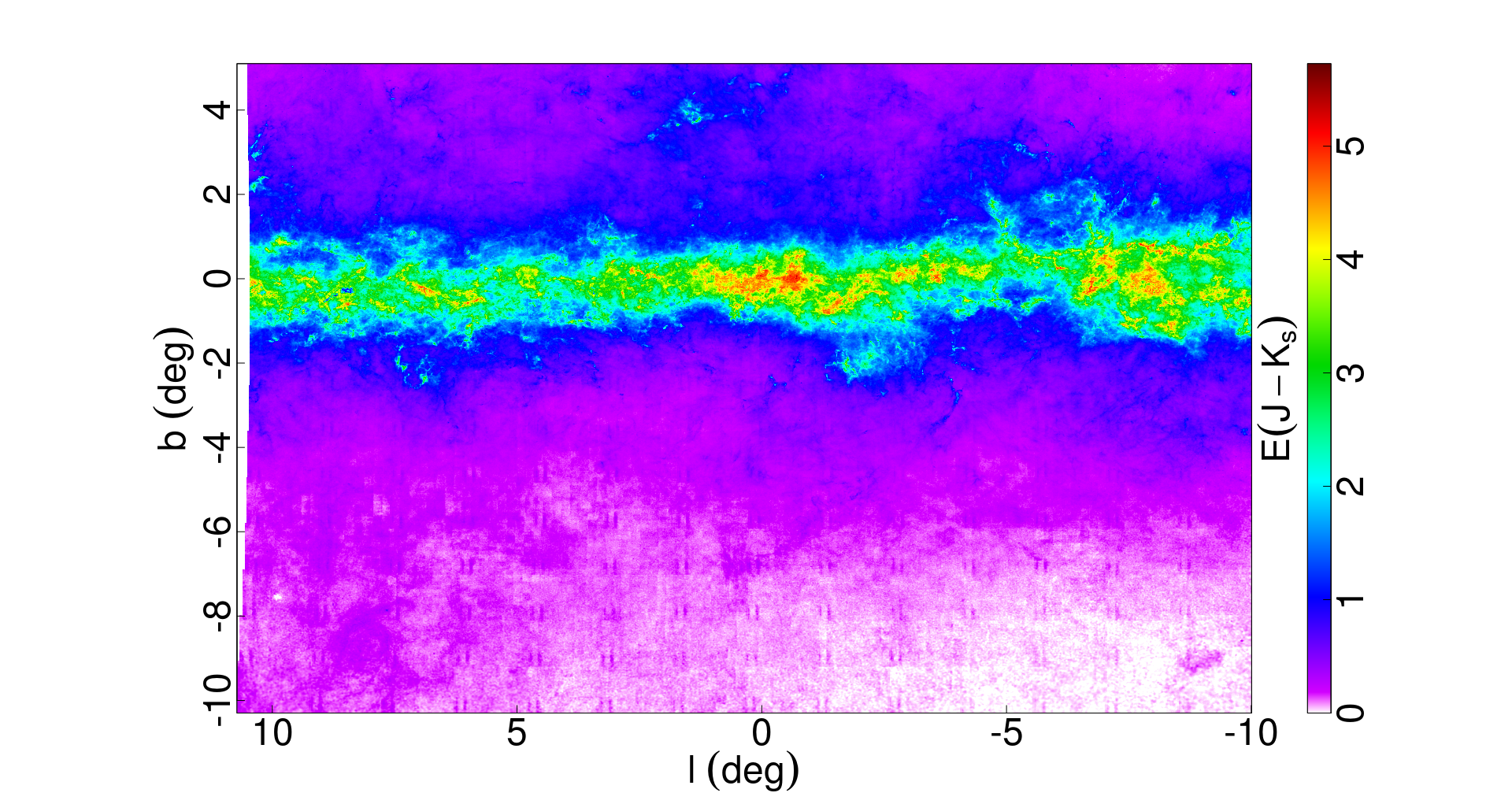}
\includegraphics[width=0.99\hsize, trim={7cm 0 6cm 0}]{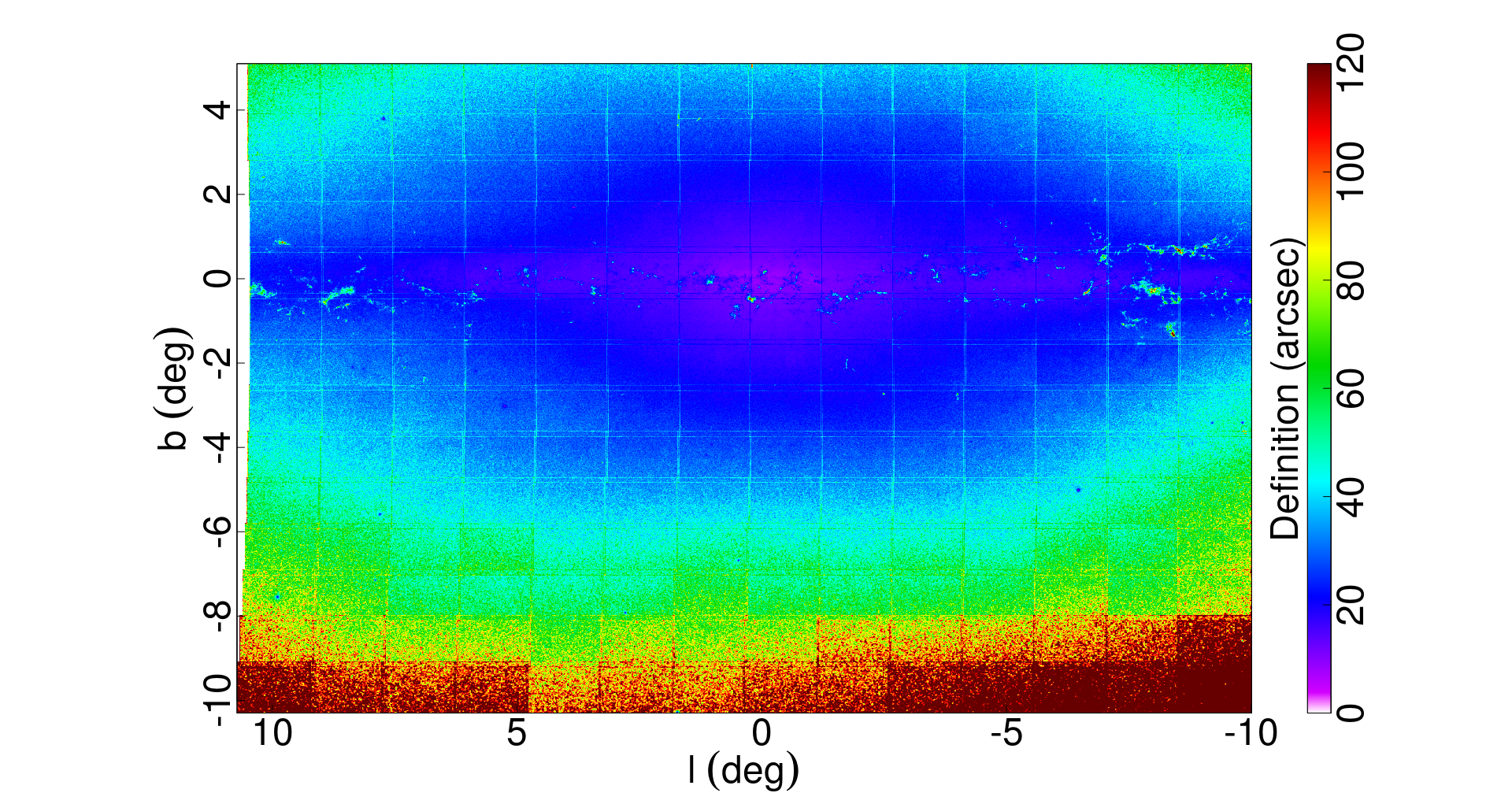}
   \caption{Top panel: Complete reddening map for the VVV bulge area. Bottom panel: Definition of the map.}
      \label{fig:wholemap}
\end{figure*}

In Fig.~\ref{fig:wholemap}, top panel, we display the whole reddening map derived in this way (it has been re-binned for plotting purposes, as the true map has a resolution of about $8540 \times 7000$\,pixels$^2$). On the bottom panel of the same figure we plotted the mean definition of the map (i.e. the average size of the structure that we can trace). The best definition is near the Galactic center, where the stellar density is highest. In theory, there we could push the resolution of the map to match its definition of $\sim\,5$\,arcsec. The outer bulge area maintains a more common $>1$\,arcmin definition.

This reddening map has been published together with the dataset from \cite{photopaper}, but it is also accessible through a dedicated site\footnote{http://basti-iac.oa-teramo.inaf.it/vvvexmap/} where users can upload a list of coordinates and retrieve the corresponding color excess values. Future additions to the map, including extensions to other Galactic regions, will be published in the same interface.

\subsection{Comparison with literature}
\label{sec:comparison}

\begin{figure*}[ht]
\centering
\includegraphics[width=0.32\hsize]{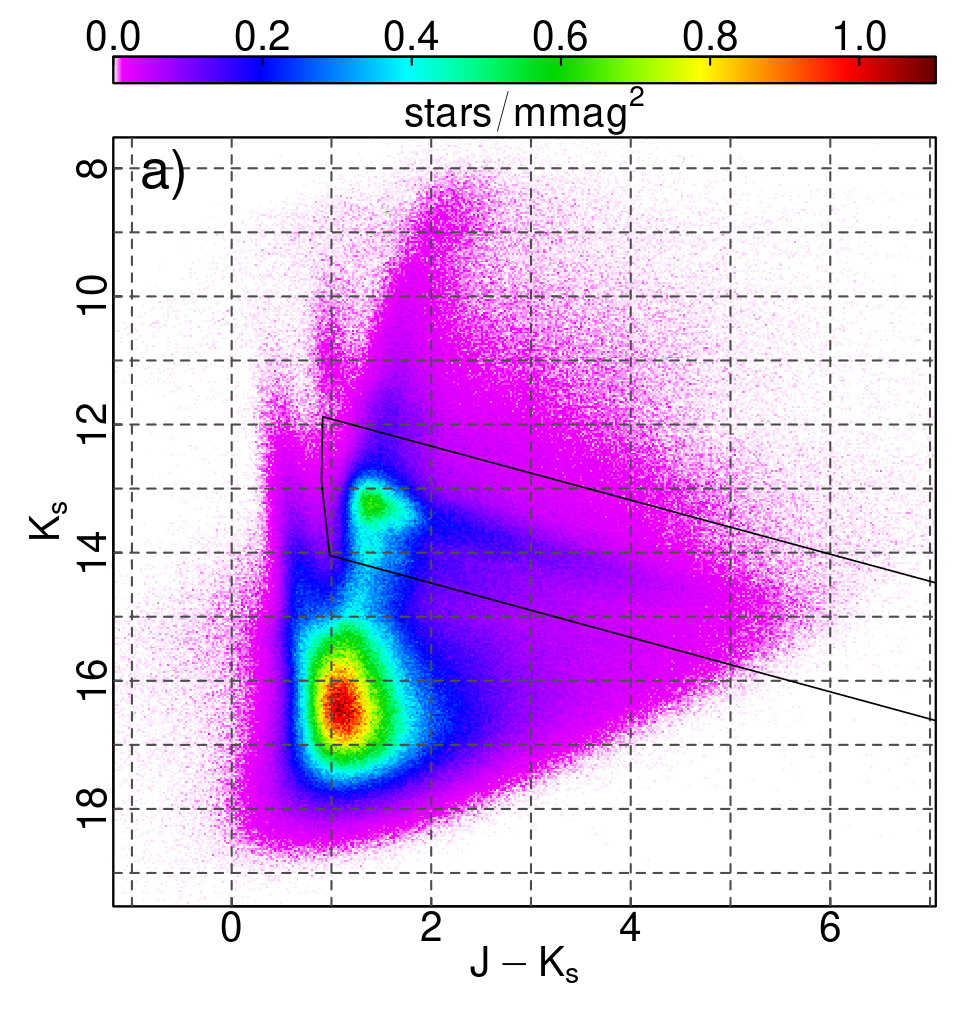}
\includegraphics[width=0.32\hsize]{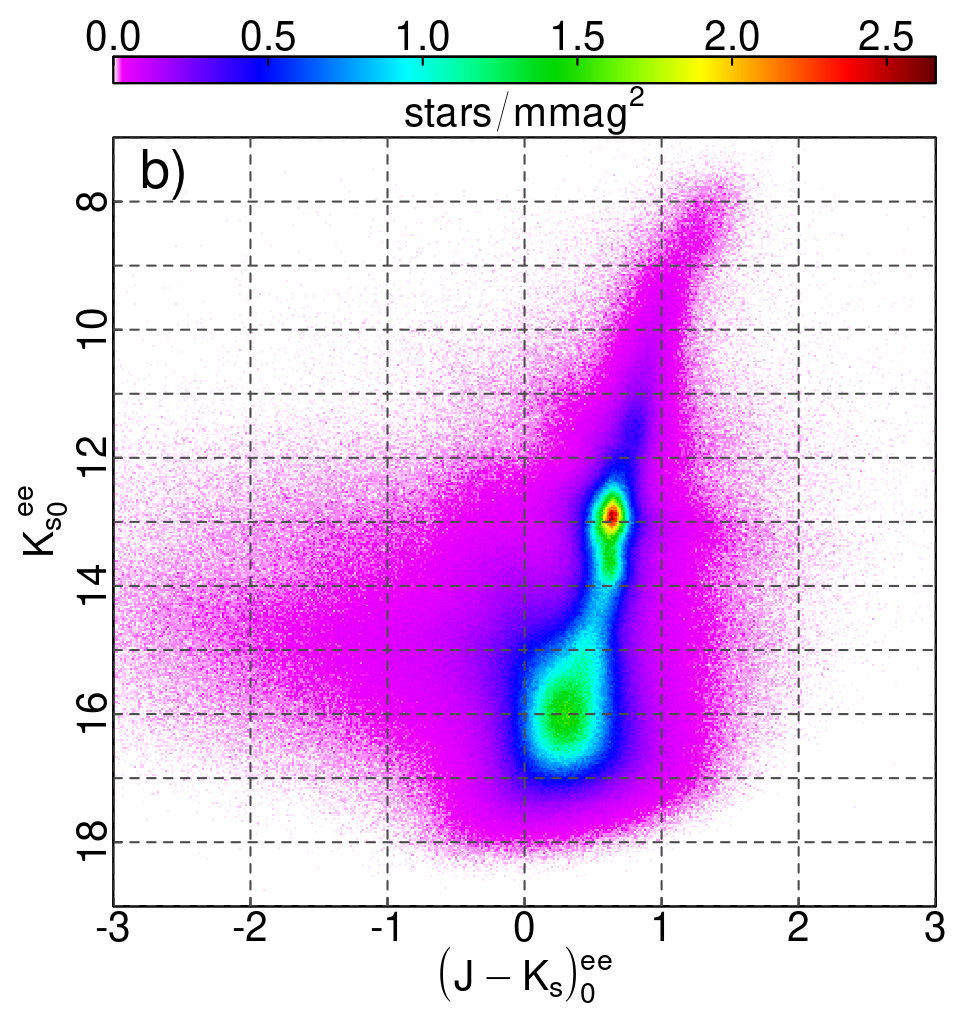}
\includegraphics[width=0.32\hsize]{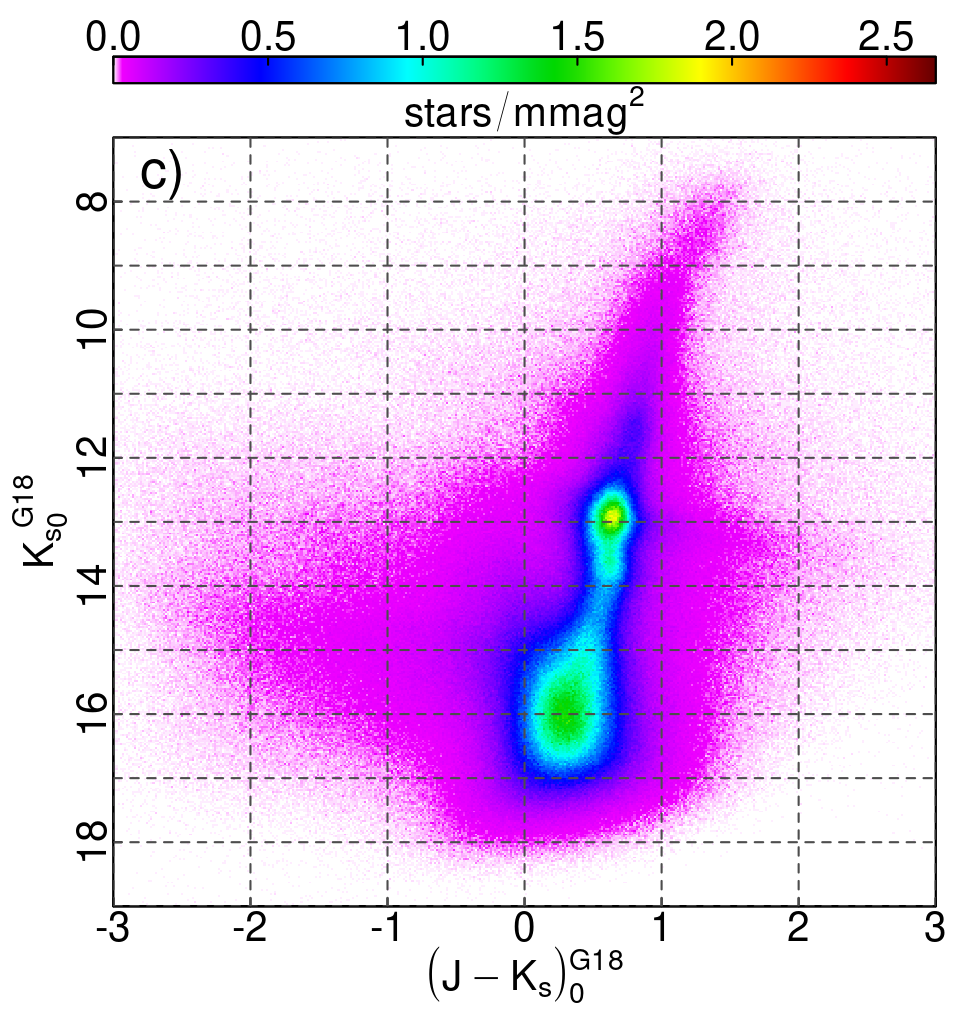}
\includegraphics[width=0.32\hsize]{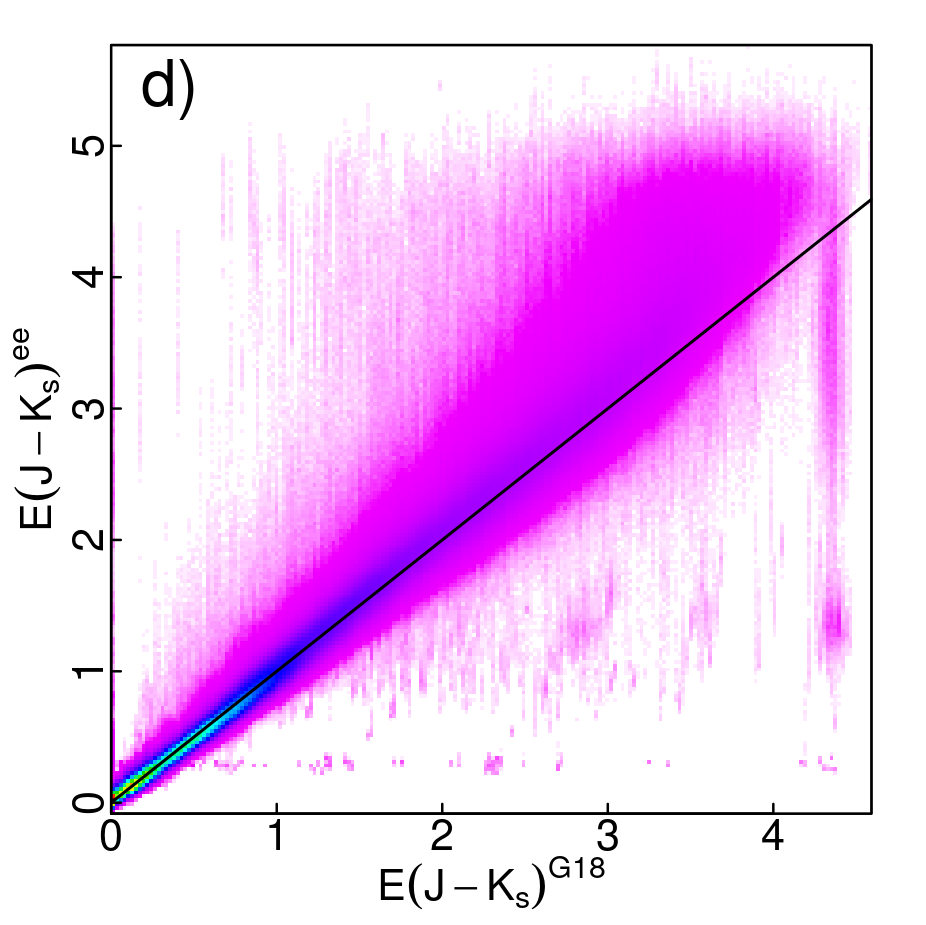}
\includegraphics[width=0.32\hsize]{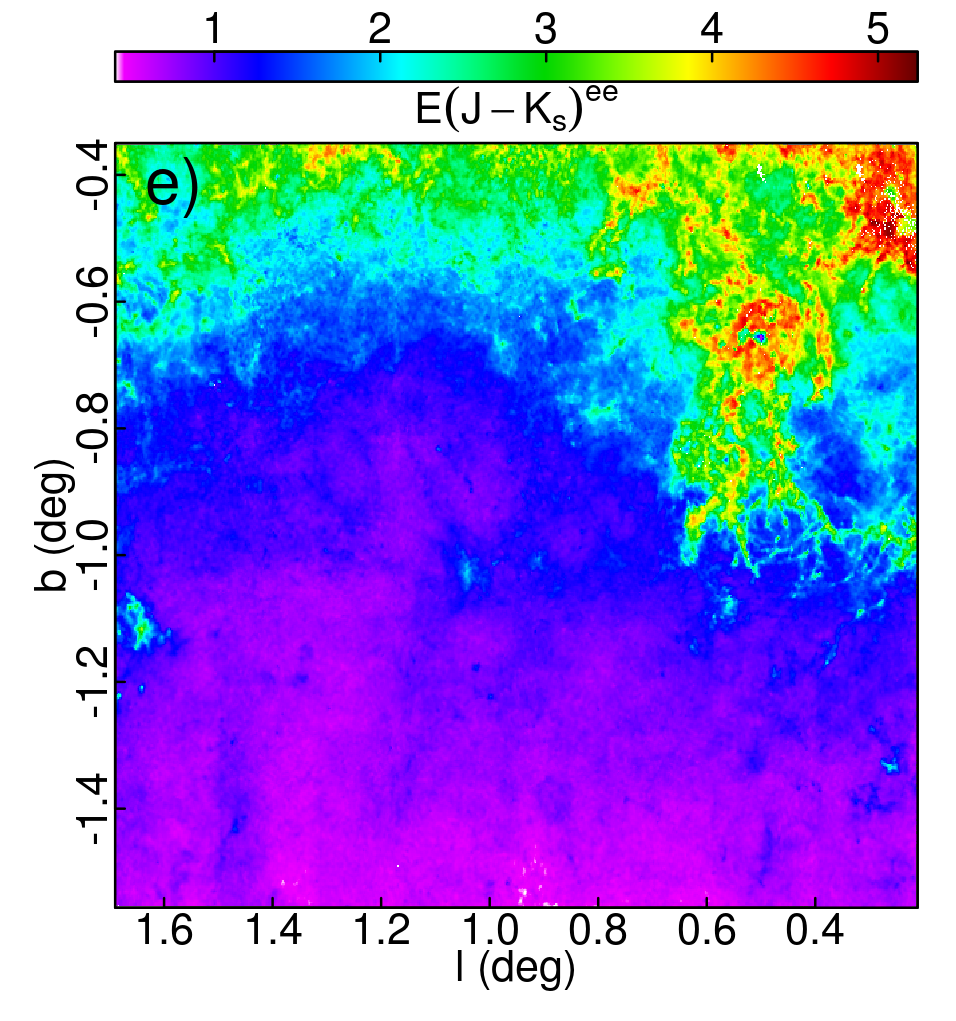}
\includegraphics[width=0.32\hsize]{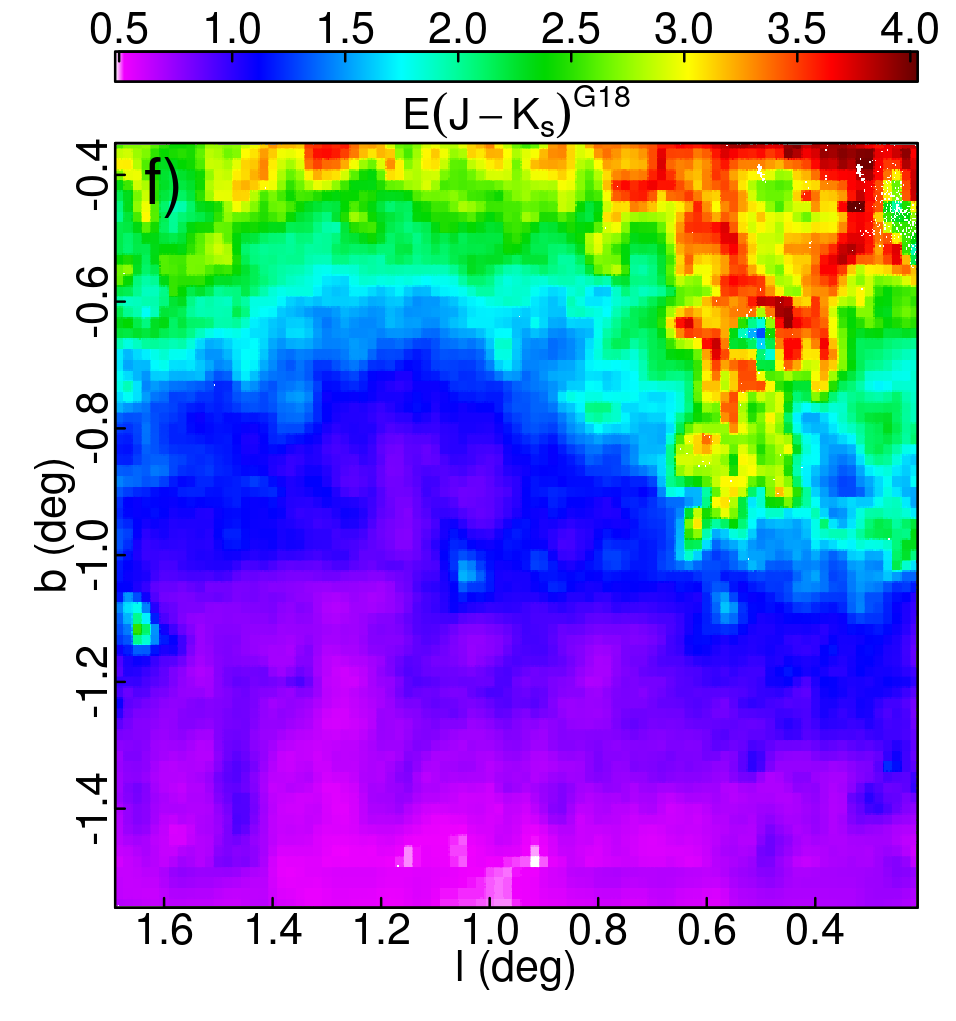}
\includegraphics[height=0.33\hsize]{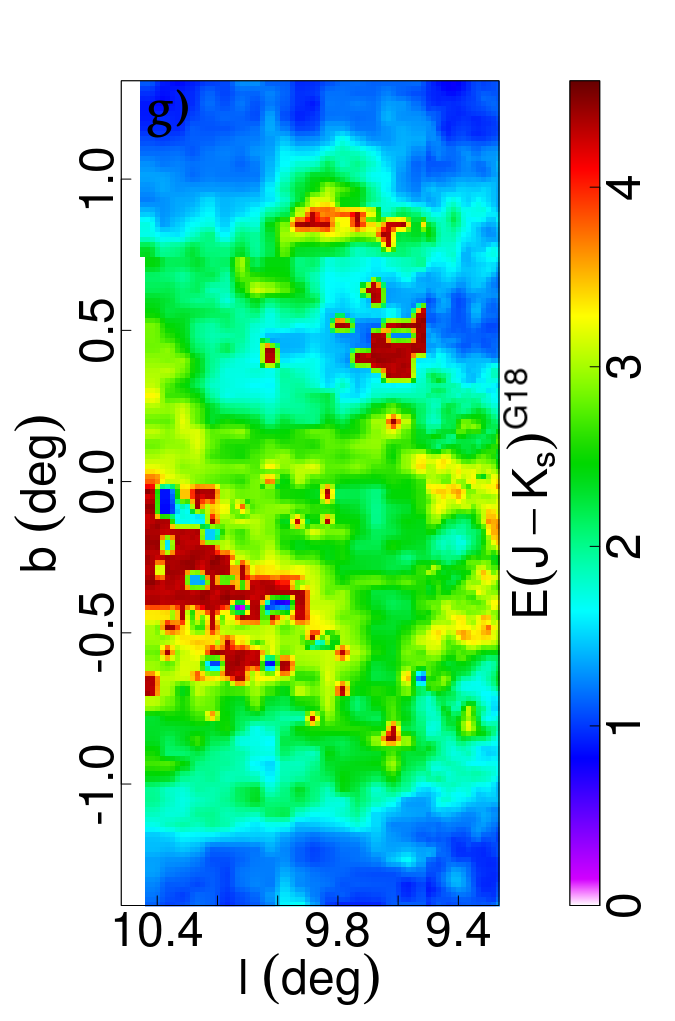}
\includegraphics[height=0.33\hsize]{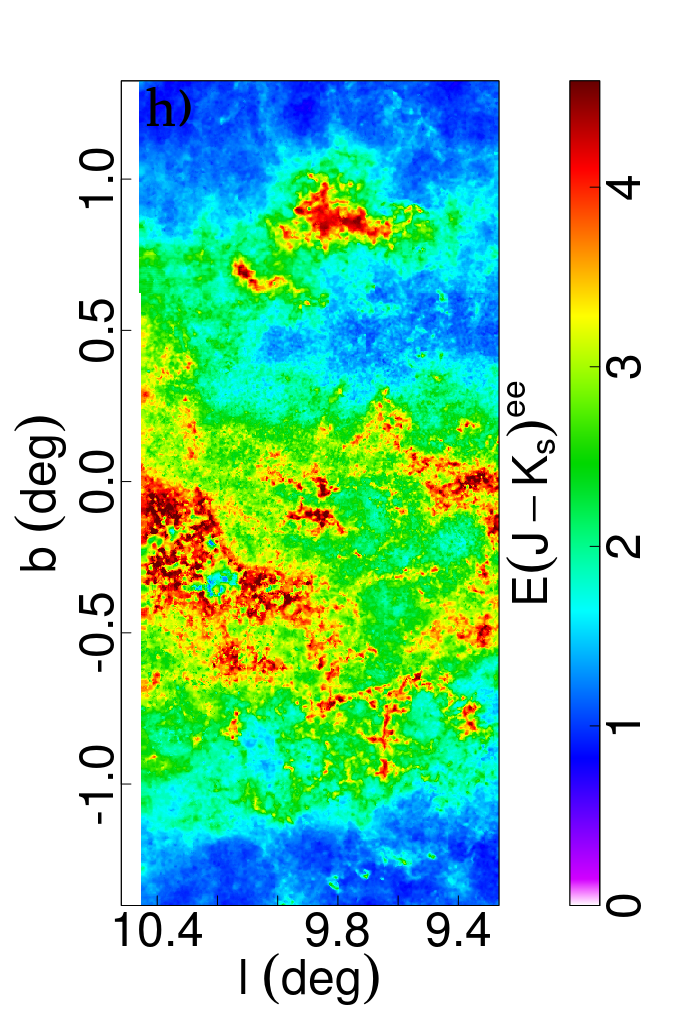}
   \caption{Comparison set of this work's map and G18. {Panel a)}: Uncorrected CMD for b320, the solid line represents the selection window used to define the RC+RGB sample, as explained in the text. {Panel b)}: Reddening corrected CMD for b320, using the map in {panel e)} (this work's). {Panel c)}: Reddening corrected CMD for b320, using the map in {panel f)} (G18). Both this panel and {panel b)} use the same reddening law (see Sec~\ref{sec:selection}) and scale in color, so it can be appreciated the way the RC in this panel is off and slanted redward. {Panel d)}: One-on-one comparison between the color excess values derived from this work (E$(\J-\Ks)^\mathrm{ee}$ in plot) with those from G18 (E$(\J-\Ks)^\mathrm{G18}$ in plot) for the whole VVV bulge area. A color-coded density map (high density in deep red, low in magenta) with a square root filter has been used to cleanly appreciate the correspondence between G18 and this work's color excess for low-reddening areas, while still portraying the upward trend where we find systematically higher $E(\J-\Ks)$ in high extinction areas. The peculiar features near maximum and minimum $E(\J-\Ks)^\mathrm{G18}$ comes from some pixels near the Galactic center, a few high-extinction concentrations scattered around the bulge area, but mostly from a few glitches in G18 map near the Galactic plane, exemplified in panel g), where a small portion of the plane appears randomly jagged with constant excess areas from the original \citet{oscar12}. Panel h) shows the same area as panel g), but with this work's realization.}
      \label{fig:b320_triad}
\end{figure*}

In order to compare this work with what is already available in the literature we use G18 as the main reference. This is preferred because i) it is a direct comparison of the E$(\J-\Ks)$ values, thus no conversion nor reddening law are necessary, and ii) the high resolution and homogeneous coverage of G18 makes it an ideal benchmark for this work. Global comparisons with other available maps \citep[e.g.,][]{marshal+00,oscar12,schultheis14,chen19}, are either already covered and contrasted against by G18, or would require projections and/or conversions to E$(\J-\Ks)$ using either passband transformations or a reddening law. Both the spatial 3D distribution of reddening agents as well as the reddening law are known to change with line-of-sight, and we are not prepared to compensate for these nor properly characterize them given our own dataset, severely limiting the useful information such comparisons would provide. 
In Fig.~\ref{fig:b320_triad} we show the uncorrected CMD for b320, our sample field, with the selection window for the RC+RGB sample used. Also shown are the extinction corrected CMD using the map from this work, and a comparison with the CMD corrected using G18 map with the coefficients from (\ref{eq:redlaw}). When using the extinction map from literature, it can be seen in this case, which is roughly representative from $|b| < 3^\circ$ fields, that this correction produces a RC that is extended redward, suggesting that it is under-corrected, while the corrected RC from this work appears compact. This key difference arises from both the higher resolution of the maps used here, and the deeper quality of the photometric dataset, thanks to which we can simply detect fainter stars (and hence, we can measure more reddened stars). The blurred blue sequence in both panels b) and c) of Fig.~\ref{fig:b320_triad} is a consequence of the over-correction for the disk stars, which mainly have a different and often much lower extinction affecting them.

Figure~\ref{fig:b320_triad}, panel d), also displays the correspondence between the color excess values we derive here and those from G18. In low-extinction areas (and incidentally high-latitude) there is a general agreement. However, in mid-extinction areas, with respect to color excess from G18, we have a slight systematic underestimation of the order of 0.01--0.05\,mag, while for high-extinction regions we obtain an expected overestimation trend with respect to G18, due to the higher completeness and resolution. The few spotty discrepancies between the excesses, as well as the prominent feature of roughly constant $\mathrm{E}(\J-\Ks)^\mathrm{G18}$, can be explained by the effect of \textit{glitches} in G18, exemplified in panel g) of Fig.~\ref{fig:b320_triad}, mostly coming from challenging areas from the original \citet{oscar12} map.

Finally in Fig.~\ref{fig:b320_triad}, panels e) and f) display the color excess values from this work and G18 side by side in b320 region. In this setup, we can appreciate the difference in resolution and definition. The former from the pixelated appearance of panel f) with respect to e), and the latter from the not-so-evident difference in color scale, where in panel e) there is an emergence of compact high-extinction regions inside the maximum extinction areas from panel f), which are otherwise not accounted for by G18. In both cases we can appreciate the richness of extinction structure that is present near the Galactic plane, and that with higher resolution and better definition, a much more complex and filamentary profile starts to emerge.

Despite the challenge mentioned in comparing other maps in literature, we can report that the best agreement with the map from \cite{schultheis14} happens with their E$(\J-\Ks)$ between 6.5 and 7.0\,kpc, where any farther comparisons yield a systematic underestimation from our part, from a median difference at 7.0--7.5\,kpc of 0.01\,mag, to 0.09\,mag at 10.0--10.5\,kpc. 

\subsection{Method limitations}
\label{sec:caveats}

There are some caveats that need to be mentioned, which minimally affect the quality and consistency of the derived reddening map.

\subsubsection{The intrinsic colors of the RC and RGB} 

The observed RC and RGB colors depends not only on the reddening, but also on the metallicity of the sampled population.  
However, the color variation as a function of the chemical content can be predicted by stellar evolutionary models.
According to the BaSTI models \citep{obasti,ubasti}, for a 10\,Gyr old population, a 1\,dex change in metallicity translates into approximately 0.2\,mag change in the $\left(\J-\Ks\right)$ color of a RC, and about 0.13\,mag for the RGB color at the same RC magnitude.

Recent spectroscopic surveys \citep[e.g.,][]{gibsIII,rojasarriagada+17} have found that overall the metallicity distribution of the bulge population is bimodal and spans a relatively broad but constant range, from  $\mathrm{[Fe/H]} \gtrsim -1$\,dex to $\mathrm{[Fe/H]}\lesssim +0.5$\,dex. 
Across the bulge area studied here, the observed field-to-field metallicity variations are due to changes in the relative fraction of these two components, metal-rich versus metal-poor stars, with the latter dominating mostly the outer regions only.
However, as evident from the photometric metallicity map provided by \citet{oscar13}, the mean metallicity across the bulge is confined within the range $-0.6 \lesssim \mathrm{[Fe/H]} \lesssim 0$.
Globally, this translates to $\lesssim\,0.07$\,mag uncertainty due to the peak-to-valley metallicity variation across the entire bulge region. However, the metallicity variations on a field-to-field scale (i.e. within 1\,deg$^2$ areas) are considerably less \citep[see][]{oscar13}, hence producing a much smaller uncertainty in RC color.

It is worth mentioning that different color-magnitude selections across the sampled VVV area may lead to changes in the tiles' zero-point, (i.e., the calibration needed to transform the extracted color information to the actual color excess E$(\J-\Ks)$). 
However, any discrepancy is taken into account by a self-consistency check based on the overlapping areas present between adjacent tiles, and then with a global zero-point correction (see Section~\ref{sec:calib}).

\subsubsection{Instrumental and format effects} 

There are two problems related to the original presentation of the data. The first is an issue that is inherited from the data images on which the photometry is based. For VIRCAM, there is a known issue with detector 16, that in practice induces a small magnitude (and color) shift in the stars that are captured on about one third of the detector. For this reddening map, the effect is ubiquitous but mostly only evident in low-reddening regions, and it appears as a regularly spaced pair of slightly higher reddening columns on the corner of each tile, aligned with the Galactic latitude axis. It is patent in the top panel of Fig.~\ref{fig:wholemap}, for $l \lesssim -7^\circ$.

The second is the window effect, which affects the border of each tile. This is simply due to the discrete and modular approach to the map construction, that is, that it is built from independent small chunks. Since the algorithm uses the $NAB = 20$ closest stars to any grid node to estimate the mean color, the pixels close to the border, rather than select stars roughly surrounding their center, will pick stars off to their sides, and farther out to fulfill the $NAB$ requirement. In practice, this means a sudden worsening of the definition, as well as an off-center estimation of the color excess. This effect can be appreciated in the bottom panel of Fig.~\ref{fig:wholemap}, where there is a clear net-like pattern of worse definition across the bulge area. This issue can be solved by using a coherent stellar catalog, so that for every small field of which the map is comprised, we can include stars slightly farther than the reach of the field, but without duplicity of stars (e.g. to build the map of a field in $|l| < 1^\circ$ and $|b| < 1^\circ$, we should include unique stars from $|l| < 1.1^\circ$ and $|b| < 1.1^\circ$ in the calculations).

\subsubsection{Sensitivity of the maps: the high- and low-ends}

Since the algorithm depends on RC stars to derive an accurate tracer of E$(\J-\Ks)$, if somewhere in the field the true extinction is high enough to drive the RC stars below the detection limit, then the color excess measurement will be (silently) erroneous. Considering the limits in \cite{photopaper} photometry and the expected color for the RC+RGB bulge sequence, we can only detect and measure excesses below $\mathrm{E}(\J-\Ks)_{max} \lesssim 5.9\,\mathrm{mag}$, with decreasing sensitivity as stellar crowding increases. Any area in the bulge whose true color excess is higher than this, will have a quasi-random estimated value in the map. There are circumstances that can either mitigate and aggravate this issue. If there is even a moderate reddening in the foreground (i.e. affecting the near disk stars) then the blue plume or the young disk MS itself can be pushed into the selection polygon and be considered in the RC+RGB sample as color excess tracers. In this scenario, an extremely high extinction area will suddenly appear as a very low extinction window. A most dramatic example of this can be observed in Fig.~\ref{fig:wholemap}, in a $16 \times 10$\,arcmin$^2$ window centered at $(l,b) \approx (8.48^\circ,-0.28^\circ)$, where it appears there are only a few bulge stars and an extincted disk sequence. Here the code reports a E$(\J-\Ks) \approx 2$, where it should be higher than 4\,mag. On the other hand, if red stars still dominate the area of high reddening, then what is more likely to happen is that the density of tracers will drastically drop, substantially worsening the definition of the map near that area but otherwise still producing a high extinction estimation of the E$(\J-\Ks)$ there, in a fashion similar to how saturation and non-linear responses affect photometric images; plateauing about the extinction threshold.

A final caveat comes from the high latitudes fields, and coincidentally, low extinction. In these regions we had to pay special attention to not include within the CMD polygon the local K- and M-dwarfs sequence \citep[mentioned in][]{photopaper,saitodemog} at $(\J-\Ks) \approx 0.8$, whose bright stars can intrude into the selection polygon. Although this sequence potentially exists in all the bulge area, the considerably low bulge-star density on these high latitude areas exacerbates the effect these otherwise rare stars have on the E$(\J-\Ks)$ estimations. In general, in this region we have an overall reduced good-to-bad stellar ratio, where a good star is simply a RC+RGB bulge star, and a bad one is neither (including artifacts).


\section{Discussion and conclusions}
\label{sec:end}

In this paper we have detailed the methodology we have used to derive a sub-arcmin resolution E$(\J-\Ks)$ map for the whole VVV bulge area, within $|l| \lesssim 10^\circ$ and $-10^\circ \lesssim b \lesssim 5^\circ$ (see Fig.~\ref{fig:wholemap}). In this map we have found agreement with G18 map (also based in VVV data and with the same filters) in low extinction areas, but found significant discrepancies in the highest extinction areas. We argue, that for these very high extinction areas, previous studies lack both the photometric completeness and resolution to both observe and identify extremely reddened stars, and to discern sub-arcmin sized very-high extinction patches. Regardless of the numerical discrepancy, we maintain excellent agreement with the general structure tracing of the reddening agents in the bulge area.

The top panel in Fig.~\ref{fig:wholemap} displays the whole extinction map, where we can appreciate how faint and expansive structures pop up from the granular background for $b < -6^\circ$, while the general trend going inward to $|b| < 4^\circ$ appears to be a superposition of threads and thin filaments, until the Galactic plane area shows a generally tangled, bubbly and clumpy distribution. With this level of detail, this map proves that if we want to study and understand the stars on the Galactic plane and center, especially bulge stars, we must consider not only prohibitive extinctions, but also, high levels of variability and structured profiles of dust and gas along the line-of-sight in small ($<1^\prime$) scales. A small and elusive property of this map and method is that they can also pick up small color variations produced by clustered stars, which are represented by very localized and sudden changes in color excess (usually bluer than the background) that highlight these objects. In practice, due to this phenomenon, foreground clusters visibly pop-up from the background extinction.

As mentioned in Section~\ref{sec:method}, we calculate the width of the color distribution of the sample used to procure the color excesses. We have selected this quantity because the color width is also a proxy for optical depth in the line-of-sight. Thanks to the small bin sizes ($\sim$\,9\,arcsec or $\sim$\,350\,pc from an 8\,kpc bulge) we can isolate large-scale line-of-sight excess variations and know, aside from a baseline dispersion of the sequence due to actual population diversity and photometric effects, that whatever scatter in color we observe must come from the dust and gas distribution itself. Additionally this color width can be taken at face value for simulations and artificial stellar population synthesis, in case that the use of the proper photometric data is either prohibitive or too expensive. That said, this same width can be used as a scaled uncertainty estimate of $\mathrm{E}(\J-\Ks)$, which will be provided in the dataset.

Also, we can take notice of the next layer of information added to the map, which is the definition. In panel b) of Fig.~\ref{fig:wholemap} we have mapped this value, to show how quickly it can go below the 1\,arcmin mark, and we can even share that in the most dense areas, this value is even $<5$\,arcsec, which means that the resolution we have adopted is in principle insufficient to properly characterize those areas. In fact, using these smaller bin sizes increases the maximum color excess values we recover, continuing the trend we have pointed out before, that with higher and higher resolutions come stronger peaks in the extinction distribution profile. However, having $<2$\,arcsec resolution maps inside this highest density area only gives incremental benefits, that need to be paired with better photometry and more information to be properly taken advantage of.

The accurate definition of the differential extinction in bulge fields is key for an adequate synthetic population analysis \citep[][and references therein]{methodpaper}. In similar works and anything relating to star formation history (SFH) reconstruction via CMD fitting, inadequate solutions for the differential extinction of the field being studied could have detrimental effects, from significantly reducing the information the fit can provide, since a color dispersion product of a mischaracterization of the extinction could be interpreted as a dispersion in metallicity and/or age, to rendering such a study impossible in all but small consistent extinction windows, which usually also implies either a low extinction area away from the plane, or a far too limited sample of the bulge stars in these sparse windows. As mentioned in Section~\ref{sec:calib}, even in relatively low extinction regions, this map reduces the dispersion in the RC magnitudes with respect to G18 by $\sim$20\%, and as presented in Fig.~\ref{fig:b320_triad}, can visibly improve the corrections of fields near the plane, most evident it the color distribution of the RC locus. For instance, the central tile b333 ($-1.25 < l <  0.23$ and $-0.46 < b < 0.75$) shows an improvement over the standard deviation from the $(\J-\Ks)_0$ distribution of its RC area, from 0.46\,mag with G18 to 0.24\,mag ($\sim$46\% reduction) with this work. This improvement in extinction determination, paired with a decontamination procedure, can make possible SFH studies on the plane, that would otherwise be inconclusive due to improperly handled extinction. In addition, having a solid determination of extinction is also important for target selection of spectroscopic surveys, which rely on photometric properties related to the desired stellar parameters (e.g. photometric effective temperature, classification, etc), and in particular, upcoming spectroscopic facilities (e.g., MOONS -- \citealt{MOONS}; 4MOST -- \citealt{4MOST}), which will be able to target thousands of stars across the bulge will benefit from a reliable and accurate extinction on the plane for this purpose.

This particular version of the map has already been used in \citet{methodpaper}, and incorporated into the MW-BULGE-PSPHOT dataset \citep{photopaper}, but it is also accessible through a dedicated site\footnote{http://basti-iac.oa-teramo.inaf.it/vvvexmap/} where users can upload a list of coordinates and retrieve the corresponding color excess values. Future additions to the map, including extensions to other Galactic regions, will be published in the same interface.

\begin{acknowledgements}
FS acknowledges financial support through the grants (AEI/FEDER, UE) AYA2017-89076-P, as well as by the Ministerio de Ciencia, Innovaci\'{o}n y Universidades (MCIU), through the State Budget and by the Consejer\'{i}a de Enconom\'{i}a, Industria, Comercio y Conocimiento of the Canary Islands Autonomous Community, through Regional Budget.
EV acknowledges the Excellence Cluster ORIGINS Funded by the Deutsche Forschungsgemeinschaft (DFG, German Research Foundation) under Germany's Excellence Strategy - EXC-2094 -- 390783311.
Support for MZ and  DM  is  provided by  the  BASAL  CATA  Center  for Astrophysics  and Associated Technologies through grant PFB-06, and the Ministry for the Economy, Development, and Tourism's Programa Iniciativa  Cient\'{i}fica Milenio through grant  IC120009, awarded to the Millenium Institute of Astrophysics (MAS). 
MZ acknowledges support from FONDECYT Regular 1191505.
DM acknowledges support from FONDECYT Regular 1170121.
SH and ES acknowledge grant 309290 form IAC and grant AYA2013-42781P from the Ministry of Economy and Competitiveness of Spain.

\end{acknowledgements}
\bibliographystyle{aa}
\bibliography{myrefs}

 \appendix

\section{Higher-order Voronoi diagrams}
\label{ap:highervor}
One way to understand a Voronoi diagram is by thinking of each polygon/cell in the tessellation, as the locus of points on the plane closest to any one particular seed/generator point. This is a quite useful way to represent properties of a field whose probes are distributed irregularly on the phase space. With this in mind, we can define tessellations of the phase space based on sets of unique combinations of $n$ seed points, that is, the cells in this partition would rather represent the loci of points closest to a particular set of $n$ seeds. This representation would then be useful whenever the field property depends on several probes, as is the case of the excess values estimated here.

\begin{figure*}
\centering
\includegraphics[width=0.49\hsize]{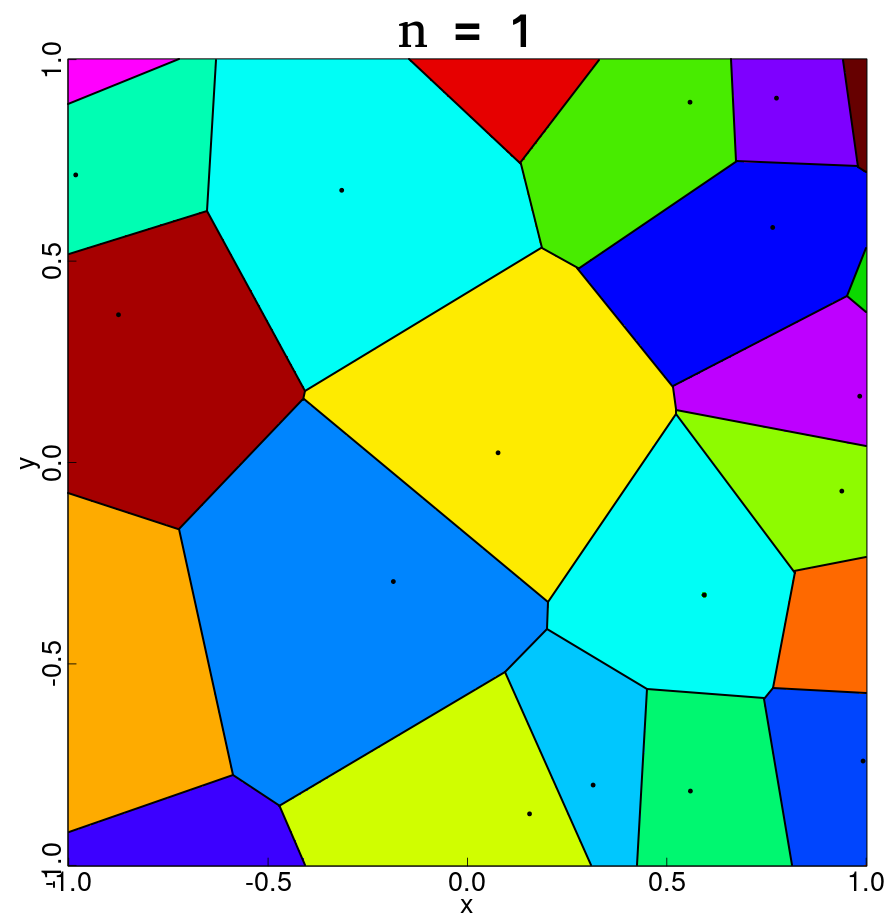}
\includegraphics[width=0.49\hsize]{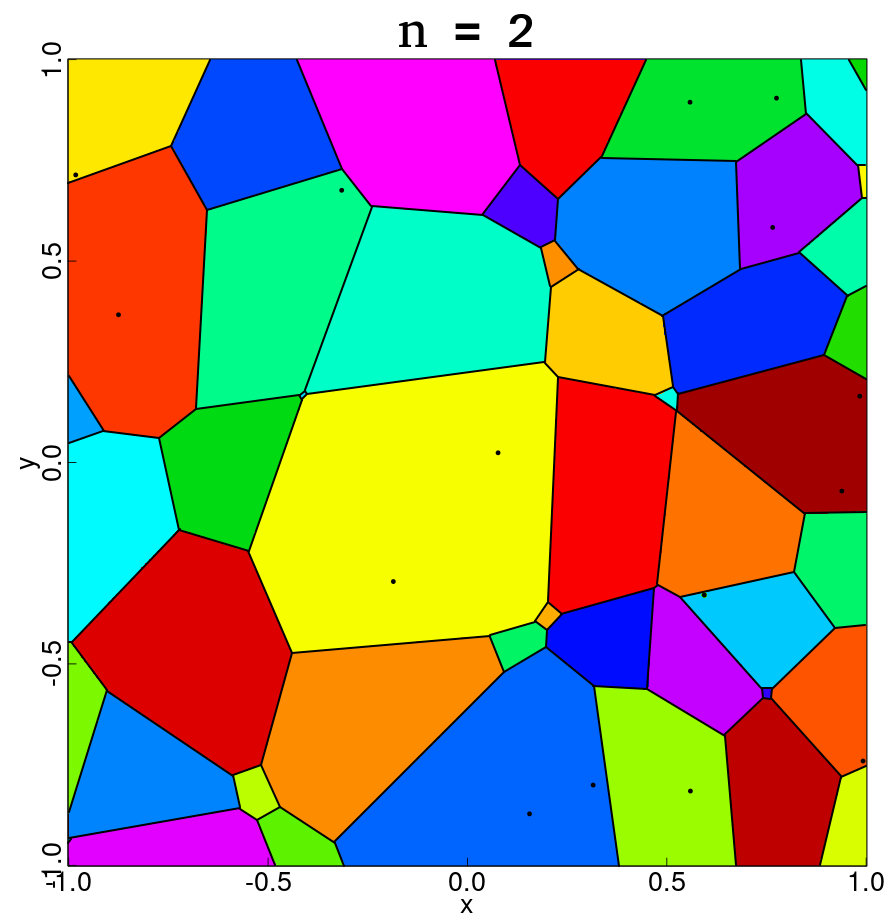}
\includegraphics[width=0.49\hsize]{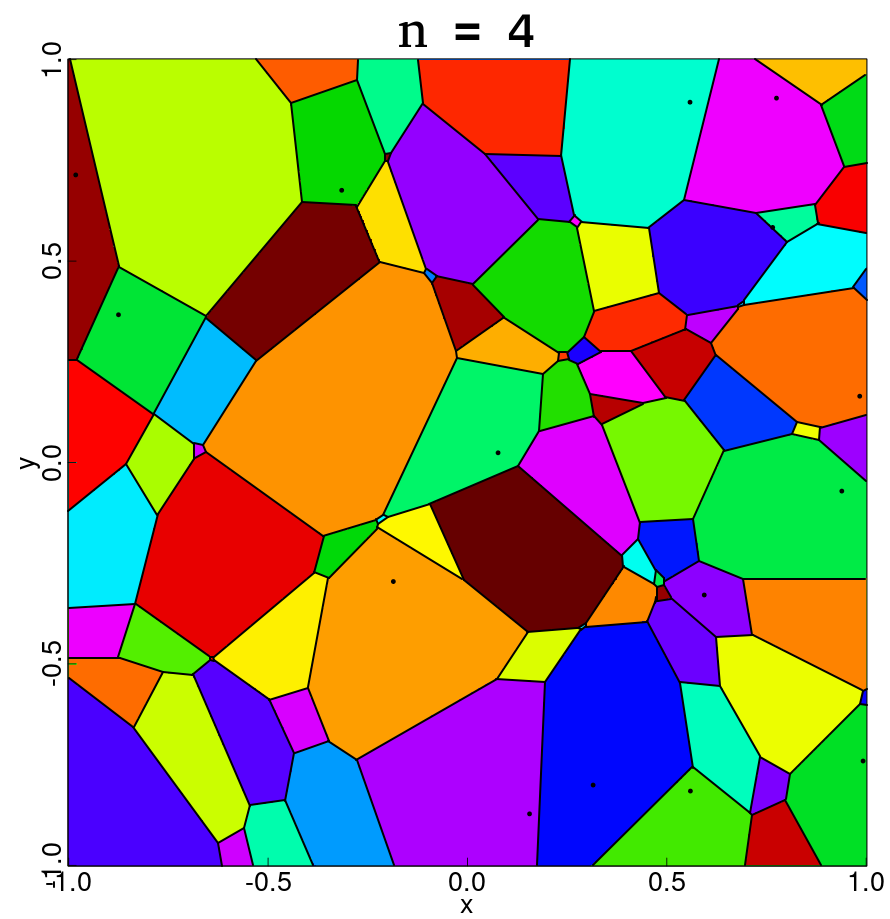}
\includegraphics[width=0.49\hsize]{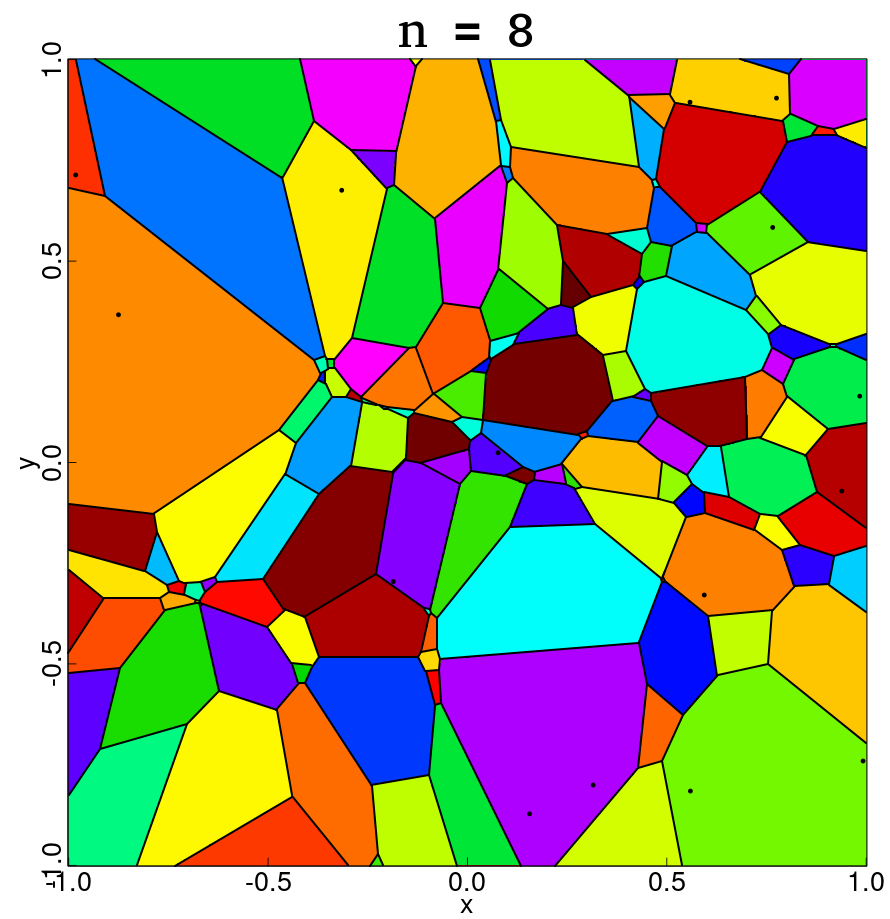}
   \caption{Voronoi diagrams of orders 1, 2, 4 and 8, of a uniformly distributed (x,y) toy seed points set where $x \in [-1.5,1.5]$ and $y \in [-1.5,1.5]$. Each panel only shows $x \in [-1,1], y \in [-1,1]$ to avoid window effects. The toy set is the same for each panel, but the order is different ($n = 1$, 2, 4 and 8, shown on top of each panel). The seed points are shown as small black circles, while each cell is randomly colored in the background and their borders delimited by solid black lines.}
      \label{fig:vortests}
\end{figure*}

In Fig.~\ref{fig:vortests}, we have depicted several Voronoi diagrams, corresponding to $n = 1$, 2, 4 and 8, for the same toy model of uniformly distributed points with $x,y \in [-1.5,1.5]$. For $n=1$ it is the familiar representation of a Dirichlet tessellation, where the seed points are all contained in their respective cells, and vice versa. However, for $n >= 2$, the polygons start to become generally smaller the higher $n$ is, with increasingly more polygons containing no seed points (and some containing more than one). While for a standard Voronoi diagram ($n = 1$) the size/area of its cells is simply a reflection of the separation of the seeds, for higher order tessellations the dimensions of these cells are generally \textit{smaller than the distance between the seeds}. In particular for this work, we have a $n = 20$ and each seed is a probe RC+RGB star. Then each tiny cell provides a uniquely defined $ee$ value, essentially defining an $ee$ field which varies in scales much smaller than typical stellar probes separations. The maps we are providing here, rather than an interpolation, are a relatively sparse evaluation of this true field, on regularly spaced nodes. In general, the these cells are bigger than the separation of the nodes, so no information is wasted. However, near the center, the probe density is so high, that even this typical node separation of $\lesssim$10\,arcsec is too big for an extensive evaluation of the field, and thus, could in principle be reduced to probe the tessellation more effectively. No particular improvement would be gained from this, however, because in these regions we appear to be limited more by depth than line-of-sight variations.

\end{document}